\documentclass[aps,prl,twocolumn,showpacs,superscriptaddress,amsmath,amssymb]{revtex4}

\usepackage{graphicx}
\usepackage{color,colordvi}
\usepackage{epstopdf}
\usepackage{dcolumn}
\usepackage{multirow}
\usepackage{color,colordvi}
\usepackage{bm}
\usepackage{psfrag}
\usepackage{subfigure}
\usepackage{amsmath}
\usepackage{ulem,colordvi,amsmath,epsfig,float,color,subfigure}
\usepackage{textcomp}%for textcelsius

\begin{document}
\title{Importance of carbon solubility and wetting properties of nickel nanoparticles for single wall nanotube growth}

\author{M. Diarra}
\affiliation{Centre Interdisciplinaire de Nanoscience de Marseille, CNRS and Aix Marseille University, Campus de Luminy, 13288  Marseille Cedex 09, France}
\affiliation{Laboratoire d'\'Etude des Microstructures, ONERA-CNRS, BP 72, 92322 Ch\^atillon cedex, France}
\author{A. Zappelli}
\affiliation{Centre Interdisciplinaire de Nanoscience de Marseille, CNRS and Aix Marseille University, Campus de Luminy, 13288  Marseille Cedex 09, France}
\author{H. Amara}
\affiliation{Laboratoire d'\'Etude des Microstructures, ONERA-CNRS, BP 72, 92322 Ch\^atillon cedex, France}
\author{F. Ducastelle}
\affiliation{Laboratoire d'\'Etude des Microstructures, ONERA-CNRS, BP 72, 92322 Ch\^atillon cedex, France}
\author{C. Bichara}
\affiliation{Centre Interdisciplinaire de Nanoscience de Marseille, CNRS and Aix Marseille University, Campus de Luminy, 13288  Marseille Cedex 09, France}

\date{\today}
\begin{abstract}
Optimized growth of Single Wall Carbon Nanotubes requires a full knowledge of the actual state of the catalyst nanoparticle and its interface with the tube. Using Tight Binding based atomistic computer simulations, we calculate carbon adsorption isotherms on nanoparticles of nickel, a typical catalyst, and show that carbon solubility increases for smaller nanoparticles that are either molten or surface molten under experimental conditions. Increasing carbon content favors the dewetting of Ni nanoparticles with respect to $sp^{2}$ carbon walls, a necessary property to limit catalyst encapsulation and deactivation. Grand Canonical Monte Carlo simulations of the growth of tube embryos show that wetting properties of the nanoparticles, controlled by carbon solubility, are of fundamental importance to enable the growth, shedding a new light on the growth mechanisms.
\end{abstract}

\pacs{61.46.Fg; 68.37.Lp}
\maketitle
The currently most widespread synthesis technique for single wall carbon nanotubes (SWNT) is the Catalytic Chemical Vapor Deposition (CCVD). Although massively developed in the last decade, the process still gives rise to unsolved questions that hamper the much coveted progress towards a selective growth of SWNTs, with designed chirality and, hence, properties. We focus here on two questions of importance in the field of nanoscience that have far reaching consequences for nanotube growth. In CCVD, for SWNT as well as graphene synthesis, a carbon rich precursor interacts with a metallic catalyst surface, possibly leading to carbon dissolution and structural modifications.

How carbon solubility and physical state of the nanoparticle (NP) depend on temperature and size is the first question we address. It has been empirically shown that efficient catalysts for SWNT growth display a limited, but non zero carbon solubility~\cite{Deck06}. Ding \textit{et al}.~\cite{Ding04} could calculate the melting temperatures of Fe-C NPs with given size and composition. Under the assumption that smaller NP size would increase the internal (Laplace) pressure, Harutyunyan \textit{et al}.~\cite{Harut08} concluded that smaller Fe-C NPs display a lower carbon solubility limit. For other transition metal-carbon systems (Ni-C, Pd-C), subsurface interstitial sites have been identified as most favorable for carbon incorporation~\cite{Klinke98, Gracia05, Yazyev08, Moors09} on flat surfaces. Assuming that this still holds true for NPs, one could expect the opposite result that carbon solubility increases for smaller NPs of these metals, because a large fraction of C should be expected close to the surface and the surface to volume ratio becomes larger for smaller sizes. In order to get a direct answer, we calculate the carbon adsorption isotherms on nickel NPs.

The second aspect is to understand how this modified carbon solubility influences the wetting and interfacial properties of nickel NPs in contact with carbon $sp^{2}$ layers. Metals are generally reported not to wet carbon nanotubes~\cite{Ebbesen96}, but capillary effects can favor the penetration of metal catalyst inside the tube~\cite{Schebar08}. In their study of the catalyst restructuring during SWNT growth, Moseler \textit{et al}.~\cite{Moseler10} assumed a 180$\char23$contact angle of Ni with graphene. As also done by Schebarchov \textit{et al}.~\cite{Schebar08} and B\"orjesson \textit{et al}.~\cite{Borj09}, the nanotube wall was considered rigid in their Molecular Dynamics simulation, and no carbon solubility was taken into account.  However, back in 1971, Naidich \textit{et al}.~\cite{Naidich71} showed that the contact angle $\theta$ of a macroscopic Ni (and also Co and Fe) drop on graphite is strongly modified by C dissolution. If this strong change of wetting properties of these typical catalysts holds true at the nanoscale, it certainly plays a major role in the growth mechanisms of SWNTs, e.g., to prevent the encapsulation of the NP by the growing tube wall. To study this, we directly simulate the structure of carbon enriched nickel NPs on a graphene layer at different compositions and temperatures.
In a last part, we show that the enhanced carbon solubility and dewetting properties of the NPs critically influence the growth of the tube wall, that is reported to have a major effect on the tube chiral distribution~\cite{Ding09, Rao12}, and therefore evidence an essential, though somewhat overlooked, property of the metal catalyst.

To perform these calculations, we use a carefully validated tight binding model to describe the interactions between nickel and carbon atoms~\cite{Amara09}. This model is included in a Monte Carlo (MC) code, working in the grand canonical (GC) ensemble, which is the natural scheme to study the incorporation of carbon in a simulation box, with given chemical potential ($\mu_{C}$) and temperature (T). The technical implementation of the model has recently been made very efficient~\cite{Los11}, enabling us to extend the range and accuracy of our previous calculations~\cite{Amara08}. To simulate CCVD, that is a surface reaction, C atoms are tentatively added or removed in a ``peel" of about ± 0.25 nm thickness from the NP surface.  As explained in~\cite{Frenkel02} this algorithm drives the system towards its equilibrium at fixed N$_{Ni}$, $\mu_{C}$ and $T$. Our model yields a melting temperature for pure bulk Ni ($T_m$(bulk) = 2010 $\pm$ 35 K) that is $\sim$15\% too high, while the relative melting temperatures of pure NPs (T$_{m}$(NP) / T$_{m}$(bulk) are 0.84, 0.81, 0.76, 0.69, 0.59 for NPs with 807, 405, 201, 147 and 55 Ni atoms, respectively~\cite{Los10}. We could also check that the maximum carbon solubility in a bulk system of 576 Ni atoms never exceeds 4-5\% in the crystalline phase, in agreement with the experimental phase diagram.
\begin{figure}
\includegraphics[width=0.93\linewidth]{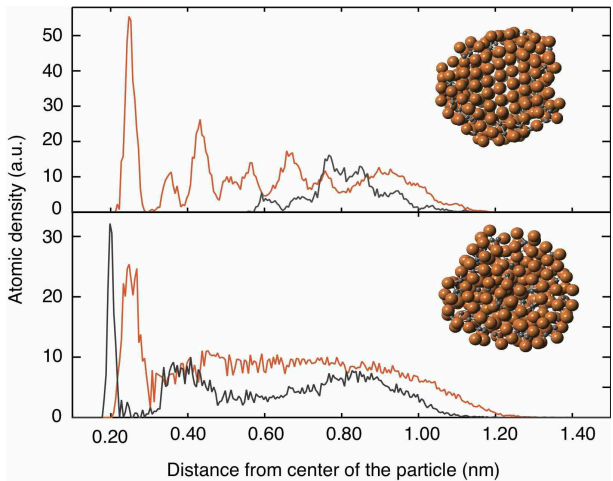}
\caption{ Structures of C-rich NPs (405 Ni; C fraction corresponding to the solubility limit; cut through the NP; Ni atoms: orange, C atoms: grey), and analysis using the distribution of radial atomic densities (black : C, red : Ni) from a central Ni atom (the closest from the NP center of mass). Bottom panel: 1400 K, particle is molten, C atoms are distributed in the whole NP. Top panel: 800 K, the core-shell structure is clearly visible, with a molten outer layer with C atoms, dimers and some trimers, while the crystalline core is pure Ni.}
\label{Figure_1}
\end{figure}

\begin{figure}
\includegraphics[width=0.93\linewidth]{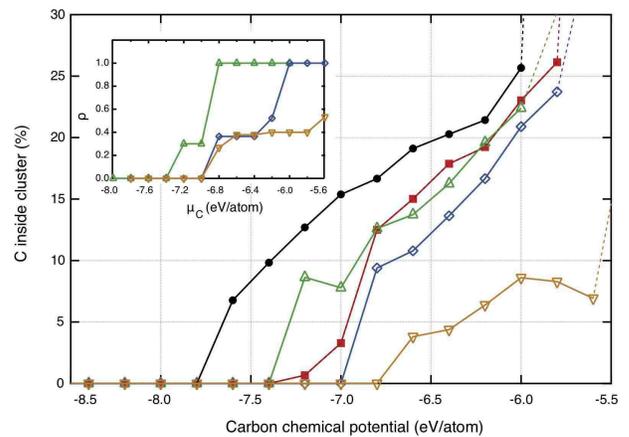}
\caption{Carbon adsorption isotherms calculated at 1000 K for nickel particle sizes 55, 147, 201, 405 and 807 atoms. 55 and 147 particles are icosahedral when pure, the others have the equilibrium Wulff shape of the finite face centered cubic structure. The y-axis is the carbon content inside the particle. Dotted lines indicate that, at the next $\mu_C$ step, carbon atoms appear on the surface. The inset presents the relative thickness of the outer liquid layer ($\rho$), as a function of $\mu_C$. NPs with 55 and 147 Ni atoms are liquid as soon as carbon is adsorbed, in these temperature and carbon chemical potential conditions.}
\label{Figure_2}
\end{figure}

How carbon solubility is modified by the nanometric size of the Ni particles can be studied by computing the carbon adsorption (or incorporation) isotherms on NPs of various sizes. We start from a bare Ni NP and run a series of GCMC simulations at different $\mu_{C}$ and $T$. Computational details are given in Supplemental Material. Once equilibrium is reached, we record the average number of C adsorbed inside and possibly outside the NP. Carbon inside the NP generally appears as dimers (C-C distance around 0.19 nm) close to the surface. This is exemplified in Fig.~\ref{Figure_1} that presents two NPs with 405 Ni, saturated with carbon, at 800 K and 1400 K. The former displays a crystalline core and a disordered outer shell with some C dissolved, while the latter is molten. Calculating the "molten" fraction is readily done using pair correlation functions. We see that the crystalline core contains few carbons, while the external molten shell is C-saturated in these examples. ``Molten" here means disordered, since, using a MC method, we have no time scale and hence no dynamics and diffusion coefficients: we cannot distinguish liquid and amorphous structures. Fig.~\ref{Figure_2} presents the average mole fraction of carbon dissolved inside NPs as a function of $\mu_{C}$ at 1000 K. Data are plotted up to the carbon solubility limit only, defined as the carbon fraction beyond which carbon species (dimers, chains ...) begin to appear on the surface of the NP. The general trend is that carbon incorporation begins at lower $\mu_{C}$ for smaller NPs. At 800 and 1000 K though, an additional effect of the NP structure can be identified. NPs with 55 and 147 are icosahedral and present a close packed surface that makes carbon adsorption or incorporation less favorable than on the (100) facets or edges of the Wulff shaped particles with 201, 405 and 807 Ni atoms. This explains why the adsorption curve for a Ni$_{147}$ icosahedral NP plotted in Fig.~\ref{Figure_2}, lies below that for a Ni$_{201}$ Wulff shaped one, as long as the NPs are not molten. Another way to read this graph is to notice that at a given carbon chemical potential, smaller particles display larger carbon content.
The incorporation of carbon modifies the physical state of the NP. At 1000 K, all particles with 55 and 147 Ni atoms are molten or amorphous, as soon as C is incorporated. Depending on $\mu_C$, and hence on their carbon content, larger particles (201 to 807 Ni) can be either molten, or display a core shell structure. The crystalline fraction of the NPs can be deduced from the inset of Fig.~\ref{Figure_2}, and a full phase diagram can in principle be computed for NPs of different sizes. Different from the conclusions of Harutyunyan \textit{et al}.~\cite{Harut08} obtained for the Fe-C system, we show here that in the case of Ni-C NPs, the solubility is larger when the particle size is smaller. This difference stems from the fact that calculations reported in Ref. \cite{Harut08} rely on the assumption that smaller size induces a larger Laplace pressure within the NP. This is true for a pure NP, but our calculations show that adsorbed C atoms strongly interact with the Ni surface and modify the surface properties of the NP, and hence its internal pressure.

In the size range (1-3 nm) studied here, that is typical for SWNTs growth, the NP surface is strongly disordered by C incorporation and fairly smooth. TEM observations, performed in situ on larger NPs~\cite{Hofmann07, Yoshida08} or after growth~\cite{Zhu05}, often show carbon walls ending on step edges of the metal catalyst. These step edges are sometimes considered as necessary for growth. Our calculations show that growth can take place without any step edge and that the movement of the tube with respect to the NP results from another driving force. \textit{In situ} Transmission Electron Microscopy (TEM) images of carbon nanofibers growth~\cite{Helveg04, Hofmann07}, as well as theoretical investigations~\cite{Schebar08, Moseler10} strongly suggest that the dewetting tendency of the NP and the tube should play a role. We therefore study the wetting of such NPs in contact with $sp^{2}$ carbon layers.

\begin{figure}
\includegraphics[width=0.93\linewidth]{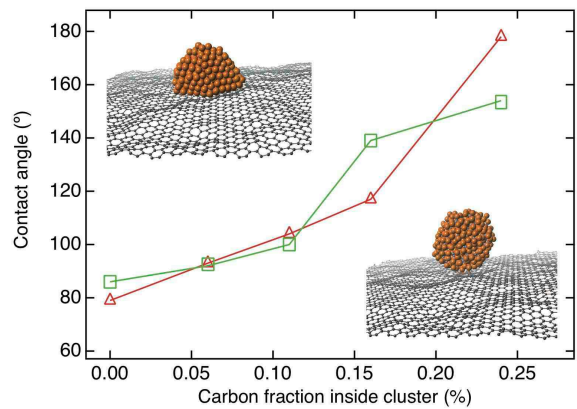}
\caption{Contact angles of a Ni$_{405}$ NP on graphene, as a function of the carbon fraction, at 1000 (red) and 1400 K (green). The two snapshots correspond to pure (left) or carbon saturated (right) NPs.}
\label{Figure_3}
\end{figure}

To separate the capillary absorption that can occur at these small nanotube sizes~\cite{Schebar08} from pure wetting, we calculate (Fig.~\ref{Figure_3}) the contact angle of NPs with 405 Ni atoms and C fractions from zero to the solubility limit, close to 25\%, on a graphene layer of 3840 C atoms, using canonical Monte Carlo simulations. A visual inspection of the resulting structures (Fig.~\ref{Figure_3}) reveals a qualitative agreement with the experimental data~\cite{Naidich71} obtained for macroscopic droplet sizes: the contact angle becomes larger and beyond 90$\char23$ when carbon fraction increases. The accuracy of our calculations is limited to about $\pm$ 15$\char23$. This results more from the non-symmetric droplets shapes (caused by their very small sizes and the atomic roughness of the substrate), than from statistical errors over the hundred configurations considered for averaging. However, the trend is clear and, either because of the nanometric size of the NP, or because of the interaction with one single graphene layer, or both, the contact angles are larger than reported in Naidich \textit{et al}.~\cite{Naidich71} for a macroscopic sample.

\begin{figure}
\includegraphics[width=0.93\linewidth]{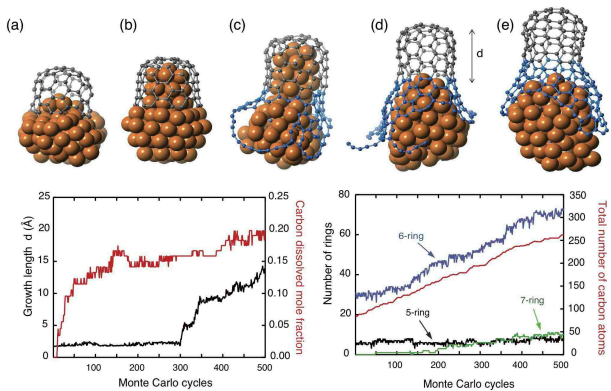}
\caption{Top panel : Snapshots of the atomic configurations during the growth of a tube butt (80 C atoms: grey) on a NP (85 Ni atoms: yellow). a) Initial configuration as prepared, b) after moderate heating and subsequent relaxation, this configuration is used to start GCMC calculations, c) MC cycle 300, before cap detachment, added C atoms are in blue, d) MC cycle 350, after cap detachment, d) MC cycle 450, during growth.
Lower panel, left: growth length d (black, defined above) and mole fraction of C dissolved in the NP (red), as a function of the Monte Carlo cycles. Right: number of carbon rings formed (blue: hexagons, black: pentagons, green: heptagons) and number of added C atoms (red), as a function of the Monte Carlo cycles.}
\label{Figure_4}
\end{figure}

In order to see how the SWNT growth mechanisms are influenced by this modification of the wetting properties of the NP when C is dissolved in, we devised the following set of computer experiments. We start by fixing short nanotube butts with about 80 C atoms on pure Ni NPs with 70 or 85 atoms. The tubes are chosen with diameters in the 0.7-0.8 nm range and different chiralities. These atomic configurations are first moderately heated and subsequently relaxed to make them reach their state of minimal energy: some Ni atoms tend to enter inside the capped tube, while remaining Ni atoms stay outside. We used them as starting configurations for a series of GCMC simulations with temperature and $\mu_{C}$ conditions chosen to allow the growth of the tubes. Out of more than a hundred trials spanning a large parameters range, we selected 10 runs for which growth was effectively promoted, for $\mu_{C}$ around -6.5 eV/atom and 1050 K $<$T$<$1300 K. For all samples, the growth starts only once the carbon content of the NP has reached about 15\%. This is slightly below the equilibrium carbon concentration values calculated above for bare NPs (around 20\%), because the C distribution within the NPs is slightly inhomogeneous, lower in the part of the NP that is protected by the carbon cap that prevents C adsorption. Attempted carbon insertions are accepted only on the free surface of the NP and, once it is saturated, C atoms diffuse on the surface to form very mobile short chains that eventually get attached to the tube lip like a garland. Growth by attachment of short carbon chains has been evidenced for graphene on metal~\cite{Loginova08}, and chain length probably depends on the experimental conditions. Fig.~\ref{Figure_4} presents a series of atomic configurations during a successful growth. As outlined in our recent combined TEM and computer simulation study~\cite{Fiawoo12}, discriminating so-called perpendicular and tangential growth modes, the most efficient one is the latter: it corresponds to the situation depicted here, with carbon walls developing along the NP's side and not pushing perpendicular to the NP surface. The growth is measured by the distance from the topmost C atom of the initial butt to its closest Ni atom and by the number of carbon rings formed, hexagons, but also defects such as pentagons or heptagons. In addition, we see the sidewall growing along the C-enriched Ni NP up to MC cycle 300. We then notice a very fast extrusion of the NP from the growing tube. Starting from a pure Ni NP, all successful growths proceed in this way. The delay between the moment when the critical carbon concentration is reached and the actual cap detachment is reminiscent of the contact angle hysteresis observed for growing or receding droplets on a surface: roughness, the amount of NP surface covered by $sp^{2}$ carbon, as well as the interaction between the tube lip and the NP, certainly plays a role. It also frequently happens, at low temperatures and high $\mu_C$ favoring C incorporation, that the cap does not detach and the growth of the side wall results in a particle encapsulation. This dewetting tendency is reversible: if we remove the dissolved C atoms, the NP tends to go back inside the tube (see Supplemental Material).

This analysis enables us to suggest that the SWNT growth results from two competing phenomena. On the one hand, once a carbon cap is nucleated on the NP that serves as a template, the amount of C dissolved in the NP favors dewetting conditions that tend to separate the growing tube from the NP. On the other hand, the carbon wall growing alongside the NP and the interaction of the NP with the tube lip prevent a complete detachment. In steady state regime, these two moves are balanced and a continuous growth is observed, with a global kinetics governed by the corresponding kinetics of these two moves, plus the feedstock decomposition and the carbon diffusion kinetics. The last two processes do not depend on the tube chirality. Both remaining processes, tube dewetting from the NP and wall growth through carbon chains attachment at the tube lip strongly depend on the tube - NP interaction and are probably a key to chiral selectivity, giving a clue to interpret recent experimental findings. We note that Motta \textit{et al}.~\cite{Motta08} report improved growth, yielding longer tubes, when sulfur is included in the feedstock, acting as a surfactant. Sundaram \textit{et al}.~\cite{Sundaram11} reported the synthesis of metallic SWNT fibers suggesting that sulfur limits the growth of the catalyst NPs. Harutyunyan \textit{et al}.~\cite{Harut09} observed a preferential growth of metallic SWNTs by modifying the ambient gas phase, and explained it by a modification of the NPs morphology induced by adsorbed species, while Chiang \textit{et al}.~\cite{Chiang09} could link the chirality distribution to the composition of their Ni$_x$Fe$_{(1-x)}$ catalyst. From our analysis, we suggest that one could also interpret these results as a modification of the interfacial properties of the metal NPs, that depend very much on their size, the temperature and the carbon chemical potential that control the carbon content.

To conclude, our findings can be summarized as follows. We first showed that in the case of nickel that presents a moderate phase separation tendency with carbon, carbon solubility tends to increase when the NP size becomes smaller. Carbon atoms are concentrated close to the surface, often as dimers. Depending on their size, carbon content and temperature, NPs can be either molten or partially molten with a crystalline core. We then showed that the increase of the contact angle with increasing C content, already noticed for a macroscopic Ni drop on graphite, is even more pronounced for a Ni NP on graphene, with contact angles varying from 80$\char23$(pure Ni) to almost 170$\char23$for C-saturated Ni. Since this C induced dewetting property has been reported for Ni, Co and Fe that are typical catalyst for SWNT growth, it is very likely that it should be included in the portfolio of necessary properties for a good catalyst. Lastly, using GCMC calculations starting from a preformed carbon tube butt fixed on a pure Ni NP, we showed that the SWNT growth takes place preferably by carbon wall growing along the NP side, the dissolved C promoting the dewetting of the NP from the tube, and also causing the NP surface to be strongly disordered (liquid or amorphous). We suggest that engineering the reactivity of the tube lip in the presence of a disordered metal, enriched with carbon, and the interfacial properties of the NP are keys to chiral selectivity.
\begin{acknowledgments}
The authors thank Drs D. Chatain, A. Loiseau and P. Wynblatt for stimulating discussions and the french research funding agency (contract ``SOS Nanotubes" ANR 09-NANO-028) for financial support.
\end{acknowledgments}

\begin{thebibliography}{0}
\expandafter\ifx\csname natexlab\endcsname\relax\def\natexlab#1{#1}\fi
\expandafter\ifx\csname bibnamefont\endcsname\relax
  \def\bibnamefont#1{#1}\fi
\expandafter\ifx\csname bibfnamefont\endcsname\relax
  \def\bibfnamefont#1{#1}\fi
\expandafter\ifx\csname citenamefont\endcsname\relax
  \def\citenamefont#1{#1}\fi
\expandafter\ifx\csname url\endcsname\relax
  \def\url#1{\texttt{#1}}\fi
\expandafter\ifx\csname urlprefix\endcsname\relax\def\urlprefix{URL }\fi
\providecommand{\bibinfo}[2]{#2}
\providecommand{\eprint}[2][]{\url{#2}}

\end{thebibliography}


\begin{references}
%
\bibitem{Deck06}
C. Deck and K. Vecchio, Carbon, 44, -2-, 267-75 (2006).
%
\bibitem{Ding04}
F. Ding, K. Bolton and A. Ros\'{e}n, J. Vac. Sci. Technol. A 22(4), 1471-6 (2004).
%
\bibitem{Harut08}
A. R. Harutyunyan, N. Awasthi, A. Jiang, W. Setyawan, E. Mora, T. Tokune, K. Bolton and S. Curtarolo,  Phys. Rev. Lett., 100, 195502 (2008).
%
\bibitem{Klinke98}
D. J. Klinke II, S. Wilke and L. J. Broadbelt, J. Catal. 178, 540-54 (1998).
%
\bibitem{Gracia05}
L. Gracia, M. Calatayud, J. Andr\`{e}s, C. Minot and M. Salmeron,  Phys. Rev. B 71, 033407 (2005).
%
\bibitem{Yazyev08}
O. Yazyev and A. Pasquarello, Phys. Rev. Lett., 100, 156102 (2008).
%
\bibitem{Moors09}
M. Moors, H. Amara, T. Visart de Bocarm\'{e}, C. Bichara, F. Ducastelle, N. Kruse and J. C. Charlier, ACS Nano, 3, 3, 511-6 (2009).
%
\bibitem{Ebbesen96}
T. W. Ebbesen, Science, 57, 95, 951-5 (1996).
%
\bibitem{Schebar08}
D. Schebarchov and S. C. Hendy, Nano Lett., 8, 8, 2253-7 (2008).
%
\bibitem{Moseler10}
  M. Moseler, F. Cervantes-Sodi, S. Hofmann, G. Csanyi, A. C. Ferrari, ACS Nano 4, 7587-95 (2010).
%
\bibitem{Borj09}
A. B\"orjesson, W. Zhu, H. Amara, C. Bichara and K. Bolton, Nano Lett. 9, 3, 1117-20 (2009).
%
\bibitem{Naidich71}
Y. V.  Naidich, V. M. Perevertailo and G. M. Nevodnik, Powder Metall. Met. Ceram. 10, 45-7 (1971).
%
\bibitem{Ding09}
F. Ding, A. Harutyunyan and B. Yakobson, Proc. Nat. Acad. Sci. USA, 106, 8, 2506-9 (2009).
%
\bibitem{Rao12}
R. Rao, D. Liptak, T. Cherukuri, B. I. Yakobson and B. Maruyama, Nature Mater. 11, 213-6 (2012).
%
\bibitem{Amara09}
H. Amara, F. Ducastelle, J.-M. Roussel, C. Bichara and J.-P. Gaspard, Phys. Rev. B 79, 014109 (2009).
%
\bibitem{Los11}
J. H. Los, C. Bichara and R. J. M. Pellenq, Phys. Rev. B, 84, 085455 (2011).
%
\bibitem{Amara08}
H. Amara, C. Bichara and F. Ducastelle, Phys. Rev. Lett., 100, 056105 (2008).
%
\bibitem{Frenkel02}
D. Frenkel and B. Smit, `` Understanding Molecular Simulation " Academic Press, London (2002).
%
\bibitem{Los10}
J. H. Los and R. J. M. Pellenq, Phys. Rev. B, 81, 064112 (2010).
%
\bibitem{Hofmann07}
S. Hofmann, R. Sharma, C. Ducati, G. Du, C. Mattevi, C. Cepek, M. Cantoro, S. Pisana, A. Parvez, F. Cervantes-Sodi, A. C. Ferrari, R. Dunin-Borkowski, S. Lizzit, L. Petaccia, A. Goldoni and J. Robertson Nano Lett., 7,3 , 602-8 (2007).
%
\bibitem{Yoshida08}
H. Yoshida, S. Takeda, T. Uchiyama, H. Kohno and Y. Homma, Nano Lett., 8, 7, 2082-6 (2008).
%
\bibitem{Zhu05}
H. Zhu, K. Suenaga, A. Hashimoto, K. Urita, K. Hata and S. Iijima, Small, 1, 12, 1180-3 (2005).
%
\bibitem{Helveg04}
S. Helveg, C. Lopez-Cartes, J. Sehested, P. L. Hansen, B. S. Clausen, J. R. Rostrup-Nielsen, F. Abild-Pedersen, J. K. Norskov, Nature 427, 426 (2004).
%
\bibitem{Loginova08}
E. Loginova, N. C. Bartelt, P. J. Feibelmann and K. F. Mc Carthy, New J. of Physics, 10, 093026 (2008).
%
\bibitem{Fiawoo12}
M.-F. Fiawoo, A.-M. Bonnot, H. Amara, C. Bichara, J. Thibault-P\'{e}nisson and A. Loiseau, Phys. Rev. Lett., 108, 195503 (2012).
%
\bibitem{Motta08}
M. S. Motta, A. Moisala, I. A. Kinloch, and A. H. Windle, J. Nanosci. Nanotechnol., 8, 2442-2449 (2008).
%
\bibitem{Sundaram11}
R. M. Sundaram, K.K.K. Koziol and A. H. Windle, Adv. Mat., 23, 43, 5064-8 (2011).
%
\bibitem{Harut09}
A. R. Harutyunyan, Gugang Chen, T. M. Paronyan, E. M. Pigos, O. A. Kuznetsov, K. Hewaparakrama, Seung Min Kim, D. Zakharov, E. A. Stach, G. U. Sumanasekera, Science, 326, 116-120 (2009).
%
\bibitem{Chiang09}
W. H. Chiang and R. M. Sankaran, Nature Mater., 8, 11, 882-6 (2009).
%

\end{references}
\end{document}